\documentclass{vgtc}                          

\graphicspath{{figures/}{pictures/}{images/}{./}} 

\usepackage{times}                     

\usepackage{tabu}                      
\usepackage{booktabs}                  
\usepackage{lipsum}                    
\usepackage{mwe}                       

\usepackage{mathptmx}                  

\newcommand{\commentout}[1]{}

\usepackage[whole]{bxcjkjatype}

\usepackage{balance}

\usepackage{fancyhdr}

\fancypagestyle{copyright}{
    \fancyhf{} 

    \fancyfoot[C]{\parbox{\textwidth}{\centering\scriptsize%
    © 2025 IEEE. Personal use of this material is permitted. Permission from IEEE must be obtained for all other uses, in any current or future media, including reprinting/republishing this material for advertising or promotional purposes, creating new collective works, for resale or redistribution to servers or lists, or reuse of any copyrighted component of this work in other works.}}
}

\onlineid{0}

\vgtccategory{Research}

\vgtcinsertpkg

\title{Development of Digital Twin Environment through Integration of Commercial Metaverse Platform and IoT Sensors of Smart Building}

\author{Yusuke Masubuchi\thanks{e-mail: y-masubuchi@kajima.com}~$^{,\ddag}$\\ %
     \scriptsize Kajima Corporation
\and Takefumi Hiraki\thanks{e-mail: t.hiraki@cluster.mu}~$^,$\thanks{These authors contributed equally to this work.}\\ %
     \parbox{1.4in}{\scriptsize \centering Cluster Metaverse Lab \\ University of Tsukuba}
\and Yuichi Hiroi\\ %
     \scriptsize Cluster Metaverse Lab
\and Masanori Ibara\\ %
     \scriptsize Cluster, Inc.
\and Kazuki Matsutani\\ %
     \scriptsize Kajima Corporation
\and Megumi Zaizen\\ %
     \scriptsize Kajima Corporation
\and Junya Morita\\ %
     \scriptsize Kajima Corporation
}

\teaser{
  \centering
  \includegraphics[width=\linewidth]{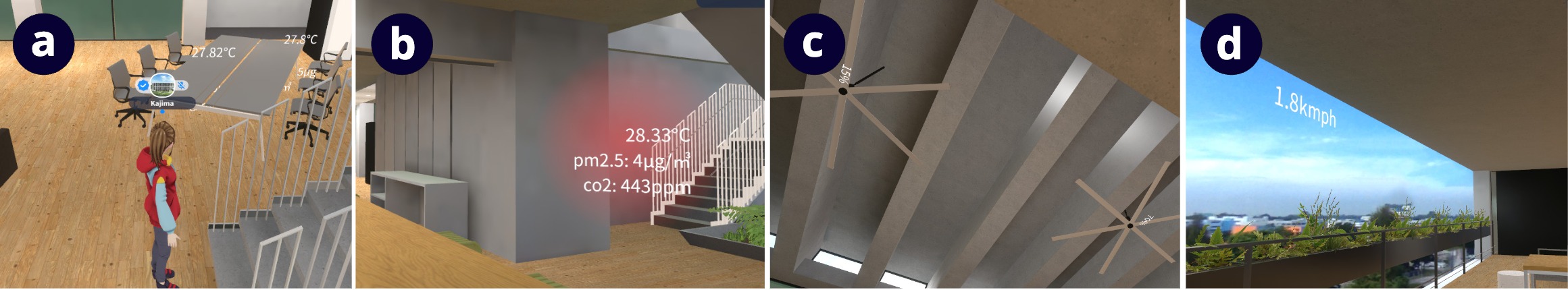}
  \caption{Environmental information visualization in a smart building digital twin environment using a commercial metaverse platform. (a) Real-time environmental data is accessible to users through a globally accessible virtual world. (b) Indoor air quality (IAQ) sensor data is visualized through color effects in the metaverse space. (c) Digital fan rotation speeds in the metaverse are synchronized with physical fan speeds measured by embedded fan sensors. (d) Wind speed data from a weather sensor modulates ambient sound effects to enhance environmental immersion.}
  \label{fig:teaser}
}

\abstract{
    The digital transformation of smart cities and workplaces requires effective integration of physical and cyber spaces, yet existing digital twin solutions remain limited in supporting real-time, multi-user collaboration. While metaverse platforms enable shared virtual experiences, they have not supported comprehensive integration of IoT sensors on physical spaces, especially for large-scale smart architectural environments.
    This paper presents a digital twin environment that integrates Kajima Corp.'s smart building facility "The GEAR" in Singapore with a commercial metaverse platform Cluster. Our system consists of three key components: a standardized IoT sensor platform, a real-time data relay system, and an environmental data visualization framework. Quantitative end-to-end latency measurements confirm the feasibility of our approach for real-world applications in large architectural spaces. The proposed framework enables new forms of collaboration that transcend spatial constraints, advancing the development of next-generation interactive environments.
} 


\begin{document}


\firstsection{Introduction}
\maketitle
\thispagestyle{copyright}

A digital twin is a technology that creates a digital replica of physical objects or processes, characterized by bi-directional synchronization, where changes in one space are automatically reflected in the other~\cite{Kritzinger2018-kj}. As physical and cyber spaces become increasingly integrated, digital twin technology plays a critical role. However, because traditional digital twin implementations have focused primarily on accurate replication of physical spaces, digital twin environments have typically been developed as stand-alone applications, limiting their technical capability for simultaneous multi-user access and widespread adoption.

In contrast, commercial metaverse platforms such as VRChat~\cite{VRChat}, Resonite~\cite{Resonite}, and Cluster~\cite{Cluster} have implemented features that allow multiple users to share experiences in real time. However, these platforms focus primarily on creating independent virtual worlds, and even in environments that replicate physical spaces, the actual exchange of information between physical and digital spaces remains significantly limited.

If we can establish digital twin environments with synchronized real and virtual information on commercial metaverse platforms, digital twin technology could evolve beyond mere physical space replication.
It could enable multiple users to seamlessly share virtual and physical spaces in real time, potentially replicating dynamic environments that include human activities and decision-making processes.
While previous research on the integration of custom IoT devices and the metaverse has demonstrated successful synchronization~\cite{Kurai2024-mr}, the integration of large-scale digital twins, such as entire architectural spaces, remains unexplored.
Bridging this technical gap could lead to the realization of Seamless Reality, which aims to integrate human perception, behavior, and cognition between physical and virtual spaces.

In this paper, we propose a digital twin environment that integrates a smart building equipped with numerous IoT sensors and a commercial metaverse platform. Specifically, we target The GEAR, Kajima Corporation's Asian R\&D center in Singapore, and reflect the office environment data collected by sensing devices in each area in real time in a virtual space built on Cluster, a multi-device compatible metaverse platform that can be accessed by smartphones, PCs, and VR-HMDs.

To achieve this, we first developed an IoT sensor platform that integrates data from the various IoT sensors installed in The GEAR and processes it in a standardized format. We then built a server system that connects this platform to Cluster, enabling real-time data transfer.
In our experiments, we measured the latency between data acquisition from IoT sensors in the physical space and its reflection in the virtual space on Cluster.
Finally, we explored methods to effectively visualize environmental information such as temperature, fan speed, and wind speed in the metaverse space (Fig.~\ref{fig:teaser}a).

The implementation of our proposed digital twin environment will allow both local smart building users and remote users to share the same space and atmosphere in the metaverse. This shared experience is expected to have positive effects on the work environment, such as improved collaboration efficiency and more dynamic discussions.


\section{System Architecture}
In this section, we describe the components of a system that realizes a digital twin environment that integrates office-scale physical and metaverse spaces.

\begin{figure}[t]
  \centering
  \includegraphics[width=\linewidth]{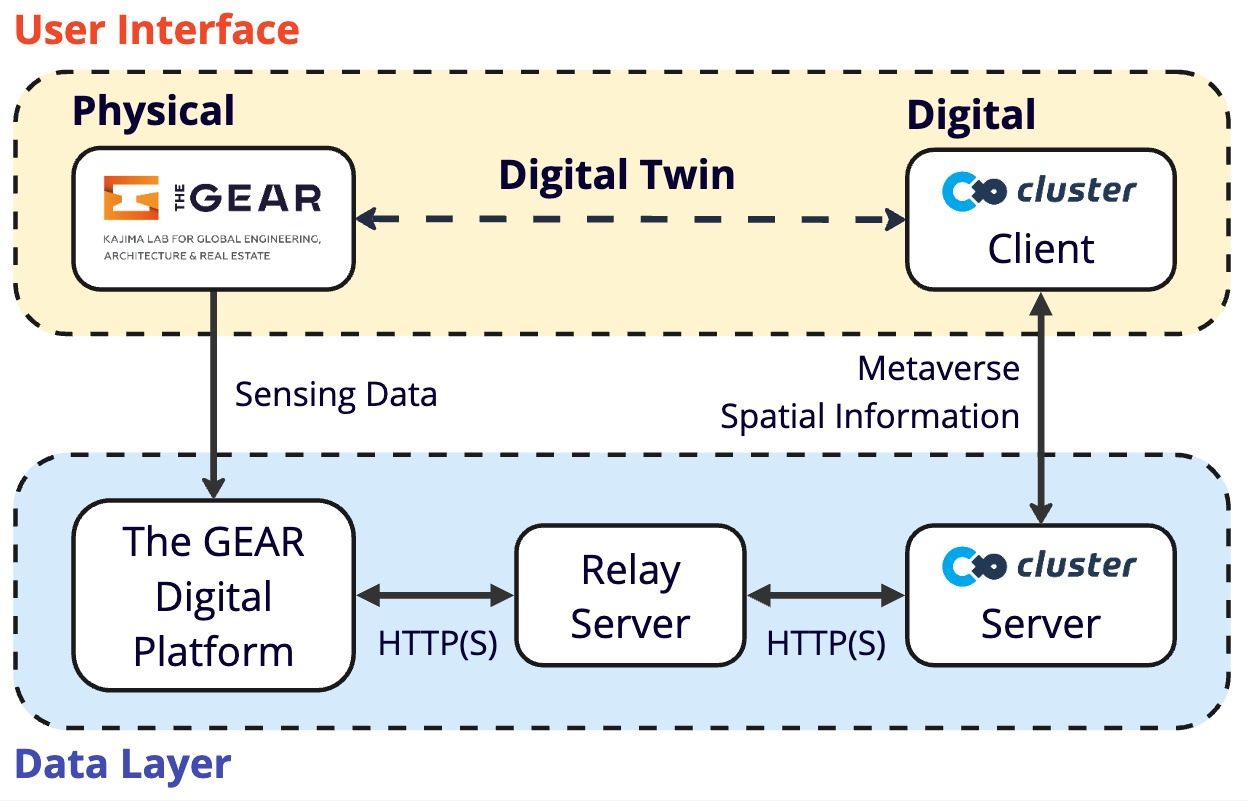}
  \caption{System architecture of the proposed digital twin environment. Sensor data from The GEAR smart building is collected by The GEAR Digital Platform (TGDPF), while spatial information in the metaverse is exchanged between the Cluster client and server. These two systems are integrated through our custom relay server that bridges TGDPF and the Cluster server infrastructure.}
  \label{fig:architecture}
\end{figure}

A diagram of the system architecture is shown in Fig.~\ref{fig:architecture}.
We utilized ``The GEAR,'' Kajima Corporation's research and development hub, as our smart building testbed equipped with IoT sensors.
The GEAR Digital Platform (TGDPF) collects and stores measurement data from the IoT sensor network deployed throughout The GEAR facility.
For the metaverse implementation, we employed Cluster as our platform and created a digital twin environment by reflecting real-time sensor measurements in The GEAR's virtual space constructed within Cluster.

To facilitate communication between systems, we developed a custom relay server that mediates information requests between TGDPF and the Cluster server.
This architecture prioritizes interoperability through its modular design: if alternative IoT sensor platforms or metaverse platforms need to be integrated in the future, only modifications to the relay server would be required. This approach ensures system flexibility and adaptability while maintaining consistent functionality.

The following subsections detail the specific implementation of each system component.


\subsection{Smart Building and IoT Sensors}

\begin{figure}[t]
  \centering
  \includegraphics[width=\linewidth]{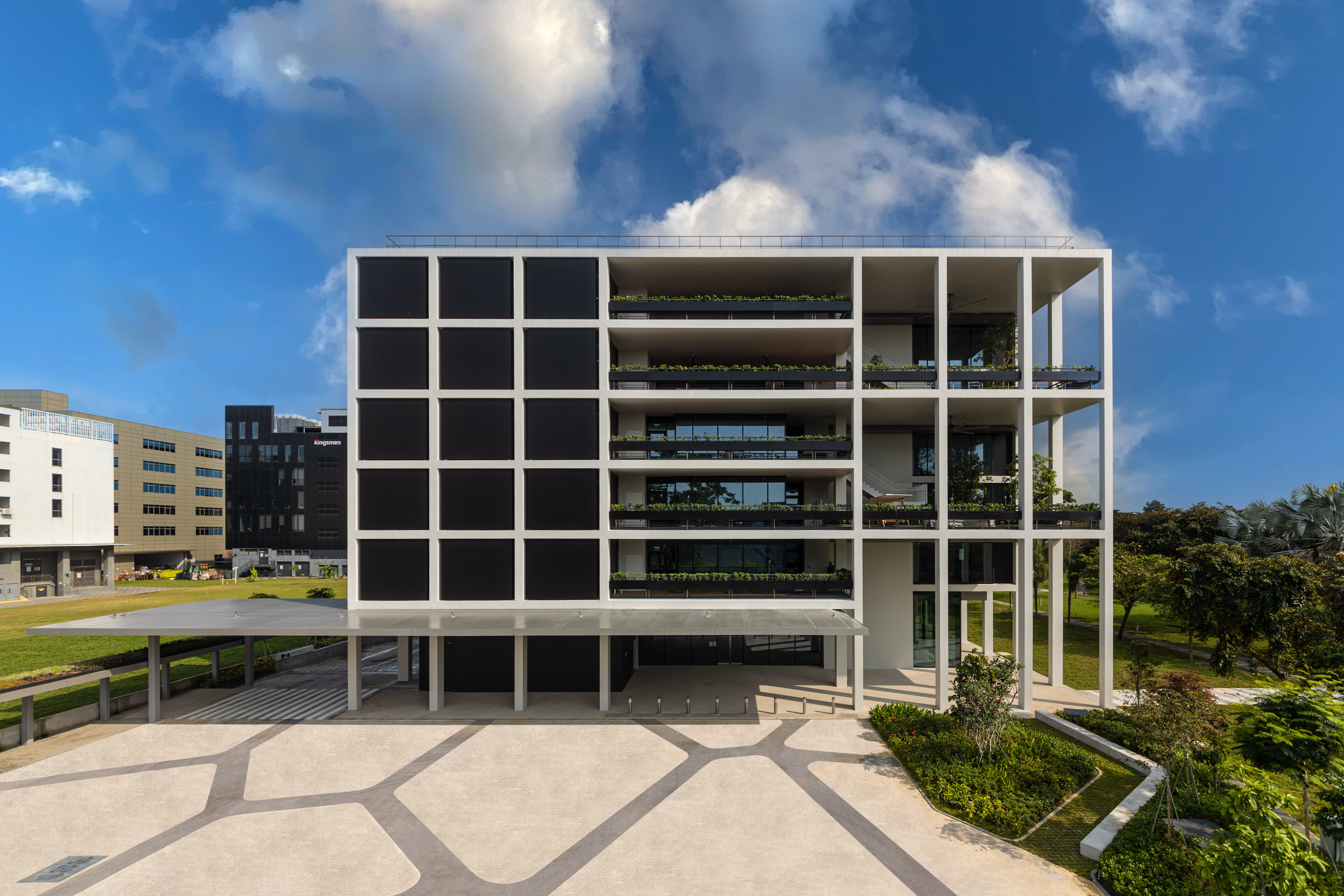}
  \caption{Appearance of The GEAR, a smart building used in our system. The GEAR is Kajima Corporation's research and development hub located in Singapore.}
  \label{fig:GEAR}
\end{figure}

\begin{figure}[t]
  \centering
  \includegraphics[width=\linewidth]{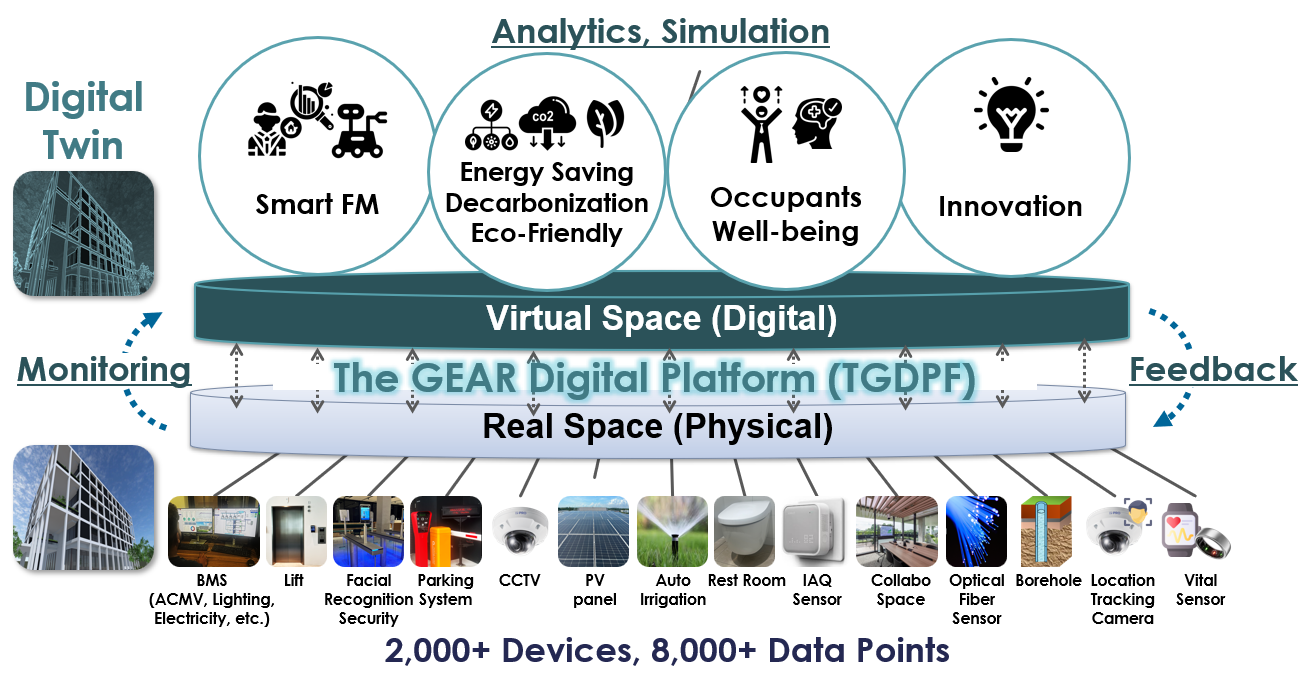}
  \caption{Overview of The GEAR Digital Platform (TGDPF). The platform integrates essential IoT functionalities into a comprehensive package, including sensor data collection, visualization, analysis, and device management capabilities. TGDPF collects and stores data from IoT sensors installed throughout The GEAR facility, enabling efficient data utilization through its API interface.}
  \label{fig:TGDPF}
\end{figure}

For our physical smart building environment, we utilized ``The GEAR'' (Kajima Lab for \textbf{G}lobal \textbf{E}ngineering, \textbf{A}rchitecture and \textbf{R}eal Estate), Kajima Corporation's research and development hub located in Singapore (Fig.~\ref{fig:GEAR}).
The facility is equipped with over 2,000 IoT sensors that quantify various indoor environmental parameters, including temperature, humidity, wind velocity, and CO2 concentration.
To centrally manage this extensive sensor network, the IoT sensor platform ``The GEAR Digital Platform (TGDPF)'' was developed.
TGDPF integrates essential IoT functionalities into a comprehensive package, encompassing sensor data collection, visualization, analysis, and device management.
The platform's API interface facilitates the effective utilization of the collected environmental data.
An overview of the TGDPF is shown in Fig.~\ref{fig:TGDPF}.


\subsection{Metaverse Platform}
For the metaverse platform, we utilize Cluster, operated by Cluster Inc.
Cluster is a commercial metaverse platform that has achieved significant market penetration, with over 1 million downloads and a cumulative user base exceeding 8 million users.
The Cluster ecosystem is comprised of two primary components: the Cluster client, which supports multiple platforms (Windows, macOS, Android, and iOS), and the Cluster server. The Cluster client transmits various user interaction data to the server, including avatar posture information, position data, voice chat audio, and emotional expressions through preset animations called ``emotes.''
The Cluster server plays a pivotal role in orchestrating message dissemination among clients while ensuring the uniform synchronization of their states. Furthermore, Cluster offers users a suite of APIs, including external communication capabilities that enable HTTP request functionality.


\subsection{Relay Server}
\begin{table*}[t]
\centering
\caption{Overview of IoT Sensor Devices Utilized in the Proposed System: Types, Functions, and Quantities.}
\begin{tabular}{|l|l|r|} \hline
Device Name & Device Function & Quantity\\ \hline
Indoor Air Quality (IAQ) Sensor & Outputs indoor air quality data including temperature, PM2.5, and CO2 concentrations & 11  \\
Fan Sensor & Monitors ceiling fan rotation speed & 4  \\
Weather Sensor & Provides weather data including wind speed and humidity & 1 \\ \hline
\end{tabular}
\label{tab:devices}
\end{table*}

\begin{table}[t]
    \centering
    \caption{Response Time for IoT Sensor Device Information Retrieval Requests.}
    \begin{tabular}{|l|r|} \hline
     Device & Response Time [ms]\\ \hline
    IAQ Sensor & 320 $\pm$ 10.6  \\ 
    Fan Sensor & 320 $\pm$ 7.27  \\ 
    Weather Sensor & 1,226 \\ \hline 
    \end{tabular}
    \label{tab:latency}
\end{table}

In this system, environmental information from ``The GEAR'' in the physical space is visualized in ``The GEAR'' in the metaverse space.
In order to realize a digital twin environment where the real and virtual are linked, it is necessary to link the API servers that manage information in the physical space and the metaverse space.
To achieve this, it is imperative to collect API data from both services, process it into a format that facilitates seamless integration, and subsequently feed it back to a designated API server.

To establish a functional bridge between these API servers, we have developed a data relay server, hereafter referred to as the relay server.The communication infrastructure between the relay server and the Cluster server and TGDPF is established using the HTTP(S) protocol, with data transmission occurring in JSON format to ensure user extensibility.
The relay server is deployed on Firebase, Google's web application development platform.


\section{Experiment}

To evaluate the effectiveness of environmental information visualization in our proposed system, we measured the data transmission latency between the Cluster client and The GEAR Digital Platform (TGDPF). We conducted 100 information retrieval requests per IoT sensor device from the Cluster client to TGDPF and measured the response times.

Table~\ref{tab:devices} presents the types, functions, and quantities of IoT sensor devices employed in our proposed system. The indoor air quality (IAQ) sensors, which output temperature, PM2.5, and CO2 concentration data, are deployed with 11 units throughout our target space. Fan sensors are embedded in ceiling fans to monitor rotation speed, with 4 units installed (matching the number of ceiling fans). A weather sensor is installed on the rooftop to provide meteorological data including wind speed and humidity.


The mean response times and standard deviations for each sensor type are presented in Table \ref{tab:latency}. Our measurements revealed mean response times of approximately 320 ms for IAQ sensors and fan sensors, and 1,200 ms for the weather sensor. The variation in response times among different sensor types can be attributed to their distinct functional characteristics.
Direct data retrieval from TGDPF, bypassing the Cluster server and relay server, yielded response times of approximately 212 ms for IAQ sensors and 215 ms for fan sensors. This suggests that routing requests through the Cluster metaverse platform may introduce an additional latency of around 100 ms.
Furthermore, the upload latency from The GEAR to TGDPF was measured across 10 trials, yielding an average of 306 ms with a standard deviation of 174 ms.
This latency can be attributed to the geographical distribution of our infrastructure, with The GEAR facility located in Singapore and TGDPF being hosted in Amazon AWS's Japan region.


Based on these measurements, our implemented data relay server demonstrates practical response times for an environmental information visualization system, achieving latencies of approximately 0.3 seconds for IAQ and fan sensors, and 1.6 seconds for a weather sensor.
These response times are well within acceptable ranges for real-time environmental data visualization applications.
Users can expect to perceive environmental information from the digital twin environment without significant delay.
Furthermore, in the most unfavorable possible situation, the total delay between a physical environmental change and its reflection in the virtual space includes both inter-server communication time and an additional latency of approximately 0.3 seconds for sensor data update. However, these combined latencies remain within acceptable bounds for practical environmental information visualization applications.
It should be noted that these response times only account for data reception at the client; additional time is required for UI rendering and other client-side processing before information display.


\section{Visualization of environmental information}

\begin{figure*}[t]
  \centering
  \includegraphics[width=\linewidth]{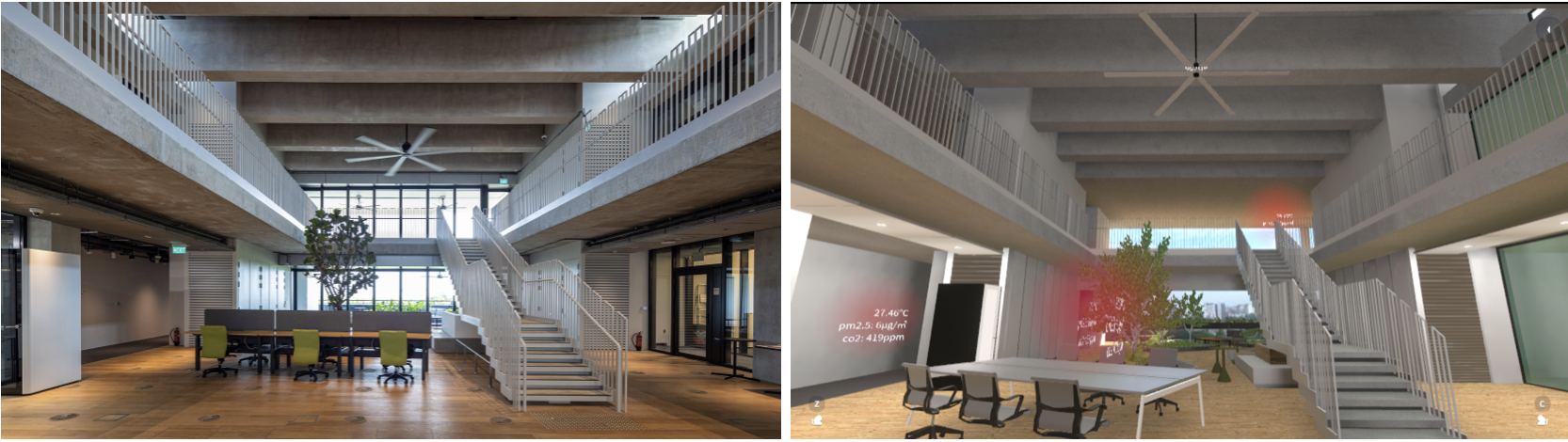}
  \caption{Comparison of the physical space (left) and metaverse virtual space (right) of K/PARK, a semi-outdoor workspace within The GEAR facility. While minor differences exist, the implementation demonstrates high-fidelity digital twin replication of the physical environment.}
  \label{fig:K/PARK}
\end{figure*}

The construction of a data relay server will make it possible to feed back environmental information from ``The GEAR'' to Cluster, the metaverse space.
Through this system, we will investigate how to effectively visualize the temperature, humidity, and wind speed of the physical office space to users in the metaverse space.

In this study, we designed a visualization method for various environmental data for the semi-outdoor workspace “K/PARK” in ``The GEAR''. A comparison of the physical office space and the metaverse space of ``K/PARK'' is shown in Fig.~\ref{fig:K/PARK}.
K/PARK takes advantage of the characteristics of Singapore's tropical climate and combines natural ventilation and ceiling fans to create a comfortable workspace that is not dependent on air conditioning equipment in principle. On the other hand, due to its characteristics, it tends to change in environmental information.


To effectively represent this characteristic environment, the following visualization techniques were employed.
First, the temperature data was represented as a change in the color tone of the space, with the intensity of the red color increasing as the temperature rose (Fig.~\ref{fig:teaser}b).
This is a method that has been commonly used in the past to visualize temperature changes in virtual spaces~\cite{Ramani2023-tp, Saeidi2021-xt}.
It is assumed that this will not only allow users in the physical space to intuitively grasp the temperature distribution, but also allow users in the metaverse space to indirectly experience the local temperature.
In addition, for air quality data such as PM2.5 and CO2 concentrations, direct display of numerical values was employed.

Next, fan speed data was provided by the fan sensor embedded in ceiling fans actually located in the ``K/PARK'' also in the metaverse space and expressing them as fan rotation speeds, providing visual feedback while maintaining the physical correspondence (Fig.~\ref{fig:teaser}c).
The ability to synchronize and coordinate the operation of hardware devices in a digital twin environment has the potential to enable a wide range of applications~\cite{Kurai2024-mr, Simiscuka2018-xr}.

For the wind speed data obtained from the weather sensor, we implemented both direct numerical display and adaptive audio feedback in the virtual space (Fig.~\ref{fig:teaser}d).
This design is based on the findings of previous research, which have shown that the acoustic effect plays an important role in recreating the sensation of wind~\cite{Giraldo2022-tw, Ito2020-sw, Ito2023-oe}.
The system dynamically changes the type of sound effects played based on the measured wind velocity values. This implementation aims to enable users to experience the local wind conditions through auditory variations, enhancing environmental awareness in the digital twin environment.


\section{Conclusion}
In this paper, we presented a shared digital twin environment that transcends geographical constraints by enabling real-time integration of environmental sensor data from ``The GEAR'' smart building into the metaverse platform Cluster.
Our custom relay server achieved practical response times, averaging approximately 0.3 s for IAQ and fan sensors, and 1.2 s for a weather sensor.
We implemented intuitive visualization methods for environmental information, including color-based temperature representation, synchronized ceiling fan rotation, and wind speed-responsive sound effects, facilitating natural comprehension of environmental conditions.
These visualization techniques enable users to intuitively understand and interact with real-time environmental data within the virtual space.

The following three points can be raised as future research issues.
First is the quantitative evaluation of the impact of environmental information visualization on user experience.
In particular, a detailed verification of the impact of different visualization methods on users' spatial perception and work efficiency is needed.
Second is the implementation and verification of bidirectional feedback from the metaverse space to the physical space. The current system is technically capable of bidirectional communication, but this has not been verified in practice.
Third is to strengthen the security of the system.
In particular, security measures for data relay servers involved in the environmental control of physical space are an important issue.

As a future prospect, the digital twin environment can be advanced by integrating more diverse sensor data, such as location data of people in addition to environmental data. If this research progresses and a digital twin environment in which multiple people can participate in real time beyond space is developed, users of smart buildings and users in remote locations can share the same space and atmosphere in the metaverse, which will lead to new possibilities, such as more efficient remote collaboration and stimulated discussion, by seamlessly linking physical space and digital space. This will create a new foundation for value creation by seamlessly linking physical and digital spaces.

\balance
\acknowledgments{
This work was partially supported by JST ASPIRE Grant Number JPMJAP2327.}

\bibliographystyle{abbrv}
\bibliography{kajima_ieeevr2025ws}

\end{document}